\documentclass[aps,pre, superscriptaddress, twocolumn]{revtex4-1}

\usepackage[dvips]{graphicx}
\usepackage{amsmath}
\usepackage{amssymb}
\usepackage{natbib}
\usepackage{color}
\usepackage{epsfig}
\usepackage{subfigure}

\begin{document}
\title{Proof of a conjecture on the infinite dimension limit  of a 
unifying model for random matrix theory}
 \author{Mario Pernici}
\email{mario.pernici@mi.infn.it}
\affiliation{ Istituto Nazionale di Fisica Nucleare, Sezione di Milano,\\ 16 Via Celoria, 20133 Milano, Italy,}
\author{Giovanni M. Cicuta}
\email{cicuta@fis.unipr.it}
\affiliation{Dip.  Fisica, Universit\`{a} di Parma,  Parco Area delle Scienze 7A, 43100 Parma, Italy}
%\date{sept.10,2018}

\begin{abstract}
We study the large $N$ limit of a sparse random block matrix ensemble.
It depends on two parameters: the average connectivity $Z$ and the size of
the blocks $d$, which is the dimension of an euclidean space.

In the limit of large $d$ with $\frac{Z}{d}$ fixed,
we prove the conjecture that
the spectral distribution of the sparse random block matrix
converges in the case of the Adjacency block matrix
to the one of the effective medium approximation, in the case of the
Laplacian block matrix  to the Marchenko-Pastur distribution.

We extend previous analytical computations of the moments of the spectral
density of the Adjacency block matrix and the Lagrangian block matrix, valid for all values of $Z$ and $d$.\\


 \end{abstract}
  
%\pacs{}

\maketitle

\section{Introduction}
This paper is a study of a sparse random matrix model in the limit of infinite dimension of the matrix. The model was presented
in \cite{CKMZ}  and it is a straightforward picture of the Hessian of a system of points connected by springs.
Its original motivation was the study of vibrational spectrum of glasses but in the present work we are not concerned with the physics insights. 
The reader is referred to \cite{CKMZ}, \cite{pari1}, \cite{pari2}, \cite{lemaitre}, \cite{scossa} and references quoted there for the usefulness of the present model in the study of a class of disordered systems. \\
The model may be seen as a generalization of the Erdos-Renyi random graph \cite{Erd}, $Z$ being the average  degree of a vertex (also called average connectivity), 
and the matrix entries being replaced by random blocks,  a set of rank-one random matrices of dimension $d$. \\

The model is studied in different regimes of the two parameters $Z$ and $d$. Certainly the role of
 the dimension $d$ of euclidean space is one relevant novelty in the classical  theory of random matrices \cite{handbook} and  the qualitative features of the model as $d$ varies from $1$ to $\infty$ agree with the expected behavior of disordered systems in spaces with different  dimension.\\

In some limiting values of the pair of parameters, several well known spectral distributions of random matrix models are obtained. Then the sparse block random matrix here analyzed interpolates among the most famous random matrix models.\\

In particular,
based on the computation of the first $5$ non-trivial moments and numerical simulations,
in \cite{CKMZ} it has been conjectured that, for $d \to \infty$
with $\frac{Z}{d}$ fixed, the spectral distribution of the
Adjacency block matrix tends
to the one of the effective medium approximation \cite{semer} and
the spectral distribution of
the Laplacian block matrix tends to the Marchenko-Pastur distribution \cite{marcpas}.
\\
An analogous result for the Laplacian matrix has been found in \cite{pari1}
in the case of regular graphs.\\

In this paper we prove these conjectures, using the following observation.
While for $d$ finite the computation of the moments reduces,
for $N \to \infty$, to contributions on walks on tree graphs, 
as in the $d=1$ case analyzed in \cite{bau},
in the limit $d \to \infty$, with $\frac{Z}{d}$ fixed, only the walks on
tree graphs
in which the sequences of edges form a noncrossing partition \cite{simi} contribute.\\

We extend the analytical computation of the moments of the spectral
distribution for the Adjacency and Laplacian $d$-dimensional block matrices
respectively to $26$th and $15$th order.  \\

After summarizing in Sect.II the definition of the matrix ensemble and the limiting domains of the parameters, we present in Sect.III the  algorithm that allows the exact automated evaluation of the several moments of the spectral distributions.\\

Sect.IV describes the proof of the convergence of this random matrix ensemble to two well known spectral distributions, in the limit of large dimension of the euclidean space. \\

In Appendix A we illustrate in detail the computation on the moments at the
first three orders.
In Appendix B we write down the moments of the spectral distribution
for the Adjacency  matrix through order $18$, and those for the Laplacian
matrix through order $10$.
In Appendix C we give some details on the distribution of the vertices in
the Laplacian matrix.  \\

It is proper to recall that in the past decade, some ensembles of block random matrices were studied, with the aim of
  capturing  modular features of complex networks.
  Possibly the most relevant model is the Stochastic Block Model, well known in the study of social and biological networks \cite{dece}, \cite{zhang}, \cite{avra}.  The model describes a complex network with $n$ nodes, partitioned into communities or blocks, often of equal size. If two nodes belong to different communities, there is an edge with probability which  depends on the chosen pair of communities. If two nodes belong to the same community there is an edge with different probability. The model is flexible enough to properly describe many nontrivial types of structures. \\
        There exists a vast and active research on networks which occasionally overlaps with random matrix theory.\\

 None of the block random matrices we know in the literature has similar properties to the sparse block random ensemble studied in this paper. However
        random block matrices were often analyzed by the  cavity method 
\cite{pari1}, \cite{pari2}, \cite{kuhn}, \cite{van}, familiar in statistical physics and this seems possible also in the present model.\\

Notation. 
We use $| v>$ and $|{\hat v}><{\hat v}|$ to indicate vectors and projectors in Dirac notation. We use $\langle x\rangle$ to indicate the ensemble expectation of a random variable $x$.\\

\section{The sparse block random matrix and the limiting domains}

We consider a real symmetric matrix $M$ of dimension $N d\times Nd$ where each row or column has $N$ random block entries, each being a $d\times d$ matrix.

The set $\{\alpha_{i,j} \}$ , $1 \leq i < j \leq N$ is a set of $N(N-1)/2$ 
i.i.d. random variables, $\alpha_{j,i}=\alpha_{i,j}$,  
with the probability law: 
$$P(\alpha)=\left(\frac{Z}{N}\right)\delta(\alpha-1)+\left(1-\frac{Z}{N}\right)\delta(\alpha)$$

In \cite{CKMZ} two prototypes of such block random matrices called the 
Adjacency block matrix $A$ and the Laplacian block matrix $L$ are studied.
In both the above matrices, the set of $X_{i,j}$, $i<j$ is a set of $N(N-1)/2$ 
independent identically distributed random matrices and each 
$X_{i,j}$ is a rank-one matrix and a projector,
$X_{i,j}=X_{j,i}=(X_{i,j})^t=|\hat a_{ij}> <\hat a_{ij}|$, 
where $|\hat{a}_{ij}>$ is a $d$-dimensional random vector of unit length,
chosen with uniform probability on the $d$-dimensional sphere.

%\begin{widetext}
\onecolumngrid

\begin{eqnarray}
A=\left( \begin{array}{cccccccc}
0 & \alpha_{1,2} X_{1,2} &\alpha_{1,3} X_{1,3}& \dots &\alpha_{1,N} X_{1,N}\\
\alpha_{2,1}X_{2,1}& 0 &\alpha_{2,3} X_{2,3} & \dots & \alpha_{2,N} X_{2,N}\\
\dots & \dots & \dots & \dots & \dots\\
\alpha_{N,1}X_{N,1} &\alpha_{N,2} X_{N,2}& \alpha_{N,3}X_{N,3}&\dots & 0 \end{array}\right)
  \qquad  \qquad
\label{d.1}
\end{eqnarray}

\begin{eqnarray}
L=\left( \begin{array}{cccccccc}
\sum_{j\neq 1}\alpha_{1,j} X_{1,j} & -\alpha_{1,2} X_{1,2} & -\alpha_{1,3}X_{1,3}& \dots & -\alpha_{1,N}X_{1,N}\\
-\alpha_{2,1}X_{2,1}& \sum_{j\neq 2}\alpha_{2,j}X_{2,j} & -\alpha_{2,3}X_{2,3} & \dots & -\alpha_{2,N}X_{2,N}\\
\dots & \dots & \dots & \dots & \dots\\
-\alpha_{N,1}X_{N,1} & -\alpha_{N,2}X_{N,2}& -\alpha_{N,3}X_{N,3}&\dots & \sum_{j\neq N}\alpha_{N,j}X_{N,j} \end{array}\right)
 \nonumber\\
\label{d.2}
\end{eqnarray}
%9
\twocolumngrid
The present sparse random block matrix may be considered a block generalization 
of the Erdos-Renyi random graph, and indeed it reduces to it for $d=1$.
It is useful to recall that, for $d=1$, the moments of the spectral 
distributions of the Adjacency matrix and Laplacian matrix were determined by 
recurrence relations at every order \cite{bau}, \cite{khor}. \\
	
In the spring model of a disordered solid, $L$ is the Hessian matrix;
the unit vector  $|\hat{a}_{ij}\rangle$ provides the direction between vertex $i$
and vertex $j$.
For more details on the Hessian matrix of disordered solids see
Refs. \cite{lemaitre}, \cite{scossa}, \cite{pari1}, \cite{pari2}.

As it is well known, any matrix elements of a power $(A^k)_{a,b}$ is the sum of
all weighted walks of $k$ steps from vertex $a$ to vertex $b$ on the $N$-vertex
graph corresponding to the matrix, where the edge $(i,j)$ of the graph has the 
weight $A_{i,j}$. We follow this traditional approach to evaluate moments of 
the matrices. In our model, $d \geq 1$, the walks to be considered in the 
$N \to \infty$ limit are the same class of walks on  trees of the simpler model
$d=1$, because of the probability law of the random variables $\{\alpha_{i,j}\}$.\\

In order to deal with the diagonal terms occurring in the Laplacian matrix, 
$L_{i,i}=\sum_{j\neq i} \alpha_{i,j}	X_{i,j}$ ,  Bauer and Golinelli, 
in the Appendix of \cite{bau},  suggested to use the property 
$\alpha_{i,j}	X_{i,j}=\alpha_{i,j}	X_{i,j}\alpha_{j,i}	X_{j,i}$. 
This extends  the walk on trees on the graph to the  case of the Laplacian 
matrix.\\
In the $d=1$ case, the average of the product of $l$ distinct weights of a 
walk is $Z^l$. The corresponding product of $j$ weights,  
in our model is $Z^l \langle tr \left(X_1 X_2\dots X_j\right)\rangle$ 
where $l$ is the number of distinct $d\times d$ random blocks, 
$1 \leq l \leq j$.   
The value of the product depends on the order of the blocks, and this prevents
recursion relations to evaluate the moments.\\
For the product of $j$ blocks, $l$ of which are distinct,
$\langle tr \left(X_1 X_2\dots X_j\right)\rangle=
\langle({\hat a}_1 \cdot{\hat a}_2)({\hat a}_2 \cdot {\hat a}_3) \dots ({\hat a}_j\cdot{\hat a}_1)\rangle$ 
is the average of the product of $j$ scalar products involving $l$ distinct 
random unit vectors.\\

Let us recall the usual definitions for the resolvent for the Adjacency and Laplacian matrices and the spectral distribution
\begin{eqnarray}
r_A(z) &=& \lim_{N\to \infty}\frac{1}{Nd}\langle Tr \frac{1}{z - A}\rangle =  \sum_{n \ge 0} z^{-n-1} \mu_n, \nonumber \\
r_L(z) &=& \lim_{N\to \infty}\frac{1}{Nd}\langle Tr \frac{1}{z - L}\rangle = \sum_{n \ge 0} z^{-n-1} \nu_n
\label{ris}
\end{eqnarray}
where $Tr$ is the trace on the $N d$-dimensional matrices and
$\mu_n$ and $\nu_n$ are the respective moments of the spectral distributions.

The spectral distribution is then evaluated by the formula
\begin{eqnarray}
 \rho(x) = -\frac{1}{\pi} \lim_{\epsilon \to 0+} \texttt{Im} \, r(x+i\epsilon)
\label{ro}
\end{eqnarray}

Three well-known random matrix models are reached by the present sparse block
random matrix in  different regimes of the parameters.
Let us briefly recall these limiting regimes.
	
\subsection{ $d=1$, the random graph}
	
  For $d = 1$ the sparse block random matrix reproduces the random graph by 
Erdos and Renyi. The spectral moments of the Adjacency matrix and the 
Lagrangian matrix are known at arbitrary order and they provide a check of our  analytic evaluations. 
More specifically, for $d=1$ we set all $c_j=1$ for all the moments of Eqs. (\ref{z.1}), (\ref{z.2}). They are polynomials in $Z$ and reproduce the moments of the random graph model. We used Table 1 and 2	in ref.\cite{bau} for the moments up to $\mu_{20}$ and $\nu_{10}$. Then we used the recursive relation (2.8) of Khorunzhy, Shcherbina, Vangerovsky \cite{khor}  to evaluate higher moments $\mu_{2n}$ and the recursive relation of Bauer and Golinelli to evaluate higher moments of $\nu_n$. (We were unable to use the recursive relation (2.9) in  \cite{khor} ).\\

%9
\twocolumngrid

	\subsection{ the dilute random graph, $Z \sim N^\alpha$ , $0< \alpha<1$}
	In this regime we let $Z$ increase with $N$. The average number of non-zero entries in each row of the random matrix is no longer finite. It is usually described as a transition from a sparse matrix to a dilute matrix.\\
	If $d=1$ several papers \cite{b1}, \cite{b2}, \cite{b3}, proved that the spectral density of the Adjacency matrix in the dilute regime is the Wigner semi-circle, and the spectral density of the Laplacian matrix is the free convolution of Wigner semi-circle with a Normal distribution.\\
	In the physics literature this was called the Addition Theorem  for two ensembles of random matrices \cite{pa1}, \cite{zee1}, \cite{jan}.
	The heuristic  explanation of the above result is obvious: in the dilute regime, the Adjacency matrix becomes an ordinary real symmetric matrix,  its spectral distribution is the Wigner semi-circle. The diagonal terms of the Laplacian are strongly dependent on the off-diagonal terms. But in the $N \to \infty$ and $Z \to \infty$ limit they are sums of a large number of i.i.d. random variables,  
so that the diagonal part of the Laplacian is made of entries which are independent
normal random variables. The Laplacian is the sum of a Wigner matrix and a diagonal of Normal variables.\\
We are not aware of analogous results for $d>1$, that is the case of random block matrices.\\
In the paper \cite{CKMZ}, it was indicated that the highest powers $Z/d$ in each moment $\mu_{2n}$ of the Adjacency matrix are the moments of the semi-circle distribution
$$\rho(x)=\frac{ \sqrt{4(Z/d)-x^2}}{2\pi (Z/d)} \quad , \quad \mu_{2n}= \frac{(2n)!}{n!(n+1)!} \left(\frac{Z}{d}\right)^n$$
It was also indicated that the two highest  powers $Z/d$ in each moment $\nu_{n}$ of the Laplacian matrix are the moments of the shifted semi-circle distribution
$$ \rho(x)=\frac{ \sqrt{8(Z/d)-(x-Z/d)^2}}{4\pi (Z/d)}$$
	This suggest that the above heuristic explanation may hold in the $d>1$ case.\\

		\subsection{ $d=\infty$ }

The third regime is the most interesting because it gives rather unexpected relations.
Let us consider the moments $\mu_{2n}$ of the Adjacency matrix, let $Z/d$ be
fixed
and  $d \to \infty$. In \cite{CKMZ} it has been verified that in this limit
the first $5$ non-trivial moments are equal to those of the effective medium
(EM) approximation by Semerjian and Cugliandolo \cite{semer}.
(It was indicated in \cite{CKMZ} how to obtain easily the moments of the effective medium approximation at arbitrary order, by Taylor expansion of the cubic equation of the resolvent).

We extend this check by computing the first $13$ non-trivial orders
(the first $9$ are written down in Appendix B).
Letting $Z/d$ fixed and  $d \to \infty$, then all the $c_j$ defined in Eq.(\ref{cdef}) tend to zero if $j > 1$, and $\mu_{2k}$ for $k \le 13$ reduce to those in the
EM approximation.

In analogous fashion, in \cite{CKMZ} it has been checked that the first
$5$ moments $\nu_{n}$ of the Laplacian matrix tend to those 
of a Marchenko-Pastur distribution. We extend this check to the first $15$
moments (the first $10$ are given in Appendix B).
These simplified moments reproduce the moments of a Marchenko-Pastur distribution, \\
	\begin{eqnarray}
	&&\rho_{MP}(\lambda)=\frac{\sqrt{(b-\lambda)(\lambda-a)}}{4\pi \lambda}\nonumber\\
	&&a=\left(\sqrt{t}-\sqrt{2}\right)^2 \quad , \quad b=\left(\sqrt{t}+\sqrt{2}\right)^2\nonumber\\
	&& t=\frac{Z}{d} \qquad 	
	\label{z.9}
  \end{eqnarray}
	The moments are all evaluated in Appendix 3 of ref.\cite{CKMZ}.\\

 \section{The moments of the spectral distributions.}
  The generating function for the moments in the Laplacian matrix is
\begin{eqnarray}
f_L(x) = \sum_{n \ge 0} x^n \nu_n &=& \lim_{N\to \infty}\frac{1}{N d}\sum_{n \ge 0} x^n \langle Tr L^n\rangle=\nonumber\\
   & =& \lim_{N\to \infty} \frac{1}{N d} \sum_{j=1}^N \langle tr \, T^{(L)}_{j}(x)\rangle
\label{muT}
\end{eqnarray}
where we define
\begin{equation}
T^{(L)}_{j}(x) = \sum_{n \ge 0} x^n  (L^n)_{j,j}
\label{T0lapl}
\end{equation}
and $tr$ is the trace on the $d \times d$ matrices.
In the Adjacency matrix we have analogous equations,
\begin{eqnarray}
f_A(x) = \sum_{n \ge 0} x^n \mu_n & =& \lim_{N\to \infty}\frac{1}{N d}\sum_{n \ge 0} x^n \langle Tr A^n\rangle=\nonumber\\
   & =& \lim_{N\to \infty} \frac{1}{N d} \sum_{j=1}^N \langle tr \, T^{(A)}_{j}(x)\rangle \quad , \quad \nonumber\\
	&&\texttt{where}\quad
 T^{(A)}_{j}(x) = \sum_{n \ge 0} x^n  (A^n)_{j,j}  \qquad
\label{muT1a}
\end{eqnarray}
 
We also evaluate the spectral density of the $d \times d$ matrix $L_{1,1}$, in the $d \to \infty$ limit.
In this case, in Eq.(\ref{T0lapl})
$(L^n)_{j,j}$ is replaced by $(L_{j,j})^n$.\\

In the large $N$ limit,  only walks 
on tree graphs contribute to the moments, in a way completely analogous to 
the Erdos-Renyi  random graph ([\onlinecite{bau}] , [\onlinecite{khor}] ). 
This can be shown in the following way.

To compute the moments, separate the indices so they are all different.
In the case of the Adjacency matrix
\begin{equation}
\sum_{j_0}(A^n)_{j_0,j_0} = \sum_{j_0,J_1,\cdots,J_{n-1}=1}^N A_{j_0,J_1} \cdots A_{J_{n-1},j_0}
\nonumber
\end{equation}
separate the indices $J_i=1,\cdots, N$ in indices $j_0,\cdots, j_{n-1}$
all different from each other, $j_r \neq j_s$ for all $r,s$.
One can do this separation
starting from the left, following the algorithm
"label and substitution algorithm" in [\onlinecite{bau}]:
the first item is labelled $0$ (index $j_0$), the next item is labelled
$1$ (index $j_1$), and so on.
Then use the rule $A_{ij} = \alpha_{i,j}X_{i,j}$.

The case of the Laplacian matrix is similar, but with the rules
\begin{equation}
L_{i,j} = -\alpha_{i,j} X_{i,j}; \qquad
L_{j,j} = \sum_J \alpha_{j,J} X_{j,J}X_{J,j}
\nonumber
\end{equation}
so that one must again separate the index $J$ occurring in $L_{j,j}$
to be a previously occurring index $j_r$ or a new index, different from
the previous ones.

One can associate a walk to a product of $X$'s thus obtained:
to the term $-\alpha_{r,s} X_{r,s}$ 
coming from the off-diagonal term one
associates the move $(r,s)$ on the graph;
the term $\alpha_{r,s} X_{r,s}X_{s,r}$ coming from the
diagonal term of $L$ can be interpreted\cite{bau} as the move $(r,s,r)$.
In the following we will use the notation $\alpha_{r,s} X_{r,s}^2$ 
for a diagonal term,
instead of simplifying it to  $\alpha_{r,s} X_{r,s}$,
when we want to emphasize that it comes from a diagonal contribution.

In the case of the Adjacency matrix there is only the move $(r,s)$.

To each move associate the variable $x$.
$(L^n)_{j,j}$ corresponds to all the walks from vertex $j$ to $j$
with $n$ moves, identified by the factor $x^n$.
The number $E$ of distinct $\alpha X$'s associated to a walk is the length
of the unoriented graph associated to the walk, leading to a factor 
$(\frac{Z}{N})^E$ after taking the averages on the $\alpha$'s.
Since the averages are on i.i.d. random variables,
after ensuring that all the sums on $V$ indices are with all different
indices, the average on the random vectors of the addends in the sums 
all give the same contribution, with multiplicity
the falling factorial $N^{\underline V} \approx N^V$ for $N \to \infty$.
Considering the extra $N^{-1}$ factor in the definition of the moment
Eq.(\ref{muT}), the contribution of this term to the moment has a factor
$N^{V-E-1}$
which for $N \to \infty$ does not vanish only for $V=E+1$, that is
only if the graph is a tree. Therefore the walk is on a tree graph.\\

In our computer implementation of the computations of moments,
after separating indices so they are all different, as described above,
and retaining only the tree graphs, one remains with the averages in
the $d$-dimensional space to be performed. In appendix A we give
in detail the computation of the first three moments in the Laplacian model;
in the appendix the terms
$\alpha_{ij}X_{ij}$ coming from the off diagonal terms of $L$, and the term
$(\alpha_{ij}X_{ij})^2$ coming from the diagonal of $L$ are kept separated.

In Appendix A of ref.\cite{CKMZ} it has been shown that the average of a power 
of a scalar product with a random unit vector ${\hat y}$ is
\begin{equation}
\langle ({\vec p}\cdot {\hat y})^{2m}\rangle_y = \frac{c_m}{d}({\vec p}\cdot {\vec p})^m
\end{equation}
where, for a unit vector $\hat a$,
\begin{eqnarray}
\frac{c_m}{d}&=&\langle\left( {\hat a}\cdot {\hat y} \right)^{2m}\rangle_y =
\frac{ (2m-1)!!}{2^m} \frac{\Gamma \left(\frac{d}{2}\right)}{\Gamma \left(m+\frac{d}{2}\right)} \quad , \label{cdef} \\
c_1&=& 1, \quad 
c_2= \frac{3}{(d+2)}, \quad c_3=\frac{5!!}{(d+2)(d+4)},  \cdots
         \nonumber
\end{eqnarray}

Taking in this equation ${\vec p} = t_1 \hat a_1 + \cdots + t_r \hat a_r$ one gets
\begin{eqnarray}
&&\langle(\hat a_1\cdot {\hat y})^{k_1} \cdots (\hat a_r\cdot {\hat y})^{k_r}\rangle_y =
\frac{k_1!\cdots k_r!}{(2m)!}\frac{c_m}{d} [t_1^{k_1}\cdots t_r^{k_r}](p^2)^m, \nonumber\\
&&\qquad k_1 + \cdots + k_r = 2m  \qquad \qquad
\label{av1r}
\end{eqnarray}
where $[M]f$ is the operation of extraction of the monomial $M$.
With this formula these averages are easily implemented in a program
using truncated products of polynomials.

For $j \neq i,k$ one has, using Eq.(\ref{av1r})
\begin{equation}
\langle <\hat a_i| X_j |\hat a_k > \rangle_{a_j} = \frac{1}{d} \left(\hat a_i\cdot \hat a_k \right)
\nonumber
\end{equation}
From this follows a simple property of the averages:
if in the trace of a product of $X$'s one of them appears only once, 
averaging on it
consists in replacing it with a factor $\frac{1}{d}$, unless it is the
only $X$ present, in which case $\langle tr X \rangle = 1$.
This property has been used, together with idempotency, to reduce the number 
of terms contributing to the moments in the computer implementation.\\

\section{The limit of large space dimension, $\frac{Z}{d}$ fixed }

In this section it will be proved that, in the limit $d \to \infty$ with $t \equiv \frac{Z}{d}$ fixed,  the resolvent of the sparse random block matrix is the resolvent of a Marchenko-Pastur  random matrix, in the case of the Laplacian matrix, or it is the resolvent of the Effective Medium Approximation by  Semerjian and Cugliangolo, in the case of the Adjacency matrix.\\

Our proof has the following steps:
Proposition 1 shows that only walks whose sequence of edges form 
a noncrossing partition are relevant to the evaluation of the spectral moments.
Proposition 2 evaluates a class of expectations which provide the limiting spectral density of the diagonal blocks of the Laplacian matrix.

After recalling the definition of primitive walks,
in Proposition 3 the generating functional corresponding to them
is written in terms of the generating functional of the moments.

Proposition 4  shows a factorization property which gives simple algebraic equations for the resolvents.\\

In the rest of this section we neglect the powers of $N$,
which always cancel for the tree graph contributions to the moments.\\

We say that a product of blocks $\prod X$ contains the pattern $abab$ if
it contains $\dots X_a \dots X_b\dots X_a\dots X_b$, with $a \neq b$.\\

\textsl{Proposition 1}.  All products of blocks, containing the pattern $abab$,
correspond to vanishing contributions to the moments of the
Adjacency matrix or the Laplacian matrix in the $d \to \infty$ limit,
$t=\frac{Z}{d}$ fixed.

Any  product of blocks $X$, where $m$ of them are distinct and which does not
contain the pattern $abab$,  contributes $t^m$ to
 the moments of the Adjacency matrix or the Laplacian
matrix in the $d \to \infty$ limit.

\textsl{Proof of the second part of Proposition 1}.
Consider a term $\prod \alpha X$ without the pattern $abab$
and with $m$ distinct blocks.
If all its $X$'s appear only once,
$\langle tr \prod \alpha X\rangle = d t^m$.
Consider a closest pair of equal $X$'s,
say $X_a...X_a$. If there is no $X$ between them, it is trivially
reduced, using idempotency. Let $X_b$ be inside the pair of $X_a$. There cannot be a second
$X_b$ inside the pair of $X_a$, otherwise the latter would not be a
closest pair; nor there can be a $X_b$ outside the $X_a$ pair, because
there cannot be the pattern $abab$. Therefore $X_b$ appears only once,
and can be replaced by $t$. The same is true for all the $X$'s between
the pair of $X_a$, which then reduce to a single $X_a$. Continuing
in the same way, all pairs can be eliminated, and one gets the contribution
$d^{-1}\langle tr \prod \alpha X\rangle = t^m$ to the moment.

\textsl{Proof of the first part of Proposition 1}.

Let us consider a product of blocks $I = \langle tr \prod \alpha X\rangle$
containing the pattern $abab$ and with $m$ distinct blocks.
Performing the averaging of the blocks occurring only once and using
idempotency as far as possible, one gets
$I = t^r \langle tr \prod^{n} \alpha X \rangle$, $ r\geq 0$, where the suffix
$n$ is the number of blocks in the reduced product and 
each block occurs at least twice.
If $h=m-r$ is the number of remaining distinct blocks in $I$, then  $n \geq 2h$.\\
The average over the $h$  unit vectors ${\hat a}_i$ has the form
 $$ \langle tr \prod^n \alpha X\rangle= Z^h  \langle \prod^{n} ({\hat a}_i\cdot {\hat a}_j)\rangle \quad , \quad i \neq j $$
%where the inner products $({\hat a}_i\cdot {\hat a}_j)$ are not necessarily
%all distinct.
From Eq.(\ref{av1r}), performing in $I$
the average over a random vector appearing $2n_1$ times, one gets a sum,
 in which each term is a constant times
$t^m d^h \frac{c_{n_1}}{d}\langle \prod^{n-2{n_1}+r_1} {(\hat a}_i.{\hat a}_j)\rangle$, with $r_1 \ge 0$.\\
Performing similarly the averages over the remaining random vectors,
one gets that
$$I = t^m d^h \sum b_{n_1,\cdots} \prod_i \frac{c_{n_i}}{d}$$
where $b_{n_1,\cdots}$ are positive rational numbers,
the number of terms in this sum is independent from $d$ and
$\sum_{i=1}^k n_i \ge \frac{n}{2} \ge h$.\\
From $\frac{c_n}{d} \le \frac{(2n-1)!!}{d^n}$ it follows that
$$0 \le \frac{I}{d} \le t^m \sum d^{h-1 - \sum_i n_i} b_{n_1,\cdots}\prod_{i=1}^h (2n_i-1)!!$$
Taking the limit of this sum for $d\to \infty$ with $t$ fixed,
the limit can be exhanged with the sum, since the latter has a finite number
of terms independent of $d$. Since $h-1 - \sum_i n_i \le -1$, it follows
that a contribution $\frac{I}{d}$ to the moment vanishes for $d\to \infty$ 
with $t = \frac{Z}{d}$ fixed.
\\

\textsl{Remark}. The noncrossing partitions \cite{simi} correspond to  the products of
blocks not containing the $abab$ pattern  \cite{klazar}.
Therefore in this limit
a walk contribution to the moment with $m$ distinct edges gives $t^m$
if its sequence of edges form a noncrossing partition, it is zero otherwise.
\\

 Let $|v>$ be an arbitrary $d$-dimensional vector, let $\{X_j\}$ be a set $N$ distinct  rank-one projectors, let $\langle..\rangle$ indicate the average over all the random unit vectors associated to the projectors $X_j$ inside the symbols $\langle..\rangle$.\\

 \textsl{Proposition 2}.
For an arbitrary $d$-dimensional vector $|v>$ one has
\begin{eqnarray}
\lim_{d\to \infty} \lim_{N \to \infty} \langle\,<v|\left(\sum_{j=1}^N \alpha_j X_j\right)^s |v>\,\rangle = <v|v>\,P_s(t)\nonumber\\
\label{p.1}
\end{eqnarray}
where the limits are done while keeping $t=\frac{Z}{d} $ fixed, $P_s(t)$ is the Narayana polynomial of degree $s$
\begin{eqnarray}
P_s(t)=\sum_{j=1}^s N(s,j)\,t^j \quad , \quad N(s,j)=\frac{1}{s}\left( \begin{array}{cc}s\\j \end{array}\right)\left( \begin{array}{cc}s\\j-1 \end{array}\right) \nonumber\\
\label{p.2}
\end{eqnarray}

\textsl{Proof of  Proposition 2}. By expanding  the power of the sum in
Eq.(\ref{p.1}), one obtains  the sum of $N^s$ terms,  each one being the product of $s$ non-commuting projectors.\\
As indicated in Proposition 1, each averaged product has a non-vanishing contribution iff it does not contain $abab$ sequences. It then contributes $t^p$ where $p$ is the number of distinct $X_j$ in the product, $1\leq p\leq s$.\\
The $s-$set of projectors $\{X_j\}$, $j=1,..,s$ is partitioned into $p$ parts, such that in  each part all projectors are equal.
The number of noncrossing partitions \cite{klazar} with $p$ parts is $N(s,p)$.\\

\textsl{Remark}. In Appendix C, it is recalled that \textsl{Proposition 2} is consequence of an old theorem by Pastur. However  the above combinatorial derivation is useful for the derivation of \textsl{Proposition 4}.\\

A primitive walk on the graph, starting  at vertex $r$, returns to it only at the last step. Our proof will use the decomposition of a generic walk into concatenated primitive walks and the generating functions corresponding to classes of primitive walks.\\
Let us define the $d \times d$ matrix $B^{(n)}_{r,s}$  to be the sum of the contributions of the primitive walks of $n$ steps, such that the first edge is $(r,s)$ and
 $B_{r,s}(x)=\sum x^n\, B^{(n)}_{r,s}$ is its  generating function.\\

Any walk starting and ending at vertex $j_0$ has
a unique representation as concatenation of primitive walks, each one starting and ending at vertex $j_0$. This implies an equation, both for the Adjacency matrix and the Laplacian matrix

\begin{eqnarray}
T_{j_0}(x) =\sum_{n\geq 0} \left( \sum_j B_{j_0,j}(x) \right)^n \qquad
\label{T00A}
\end{eqnarray}
\\

In the case of the Adjacency matrix, the walks in $B^{(A)}_{r,s}(x)$
start by definition with the edge $(r,s)$ and end with $(s,r)$;
in between there are all possible tree walks with root $s$,
generated by $T^{(A)}_{s}(x)$, which is isomorphic to $T^{(A)}_{j}(x)$ in Eq.(\ref{muT1a}).\\
 Therefore for the Adjacency matrix
the generating function of the tree primitive walks with
first edge in $(r,s)$ is
\begin{equation}
B^{(A)}_{r,s}(x) = x^2 \alpha_{r,s} X_{r,s}T^{(A)}_{s}(x) X_{s,r}
\label{Aadj}
\end{equation}

In the Laplacian matrix, the primitive walks starting with $(r,s)$
can be either the single-edge walk corresponding to $X_{r,s}^2$, or can
start with $X_{r,s}$; in the latter case, it can continue with
any tree walk rooted in $s$ and not going to $r$ (including the trivial walk),
then it can either return to $r$ with $X_{s,r}$ ending the primitive walk,
or have an edge corresponding to $X_{s,r}^2$;
in the latter case it can continue with
any tree walk rooted in $s$ and not going to $r$, and so on, so that
\begin{equation}
B^{(L)}_{r,s}(x) = \alpha_{r,s}\big(x X_{r,s}^2 + x^2 X_{r,s}\hat T_{s}^{(r)}(x)X_{s,r}\big)
\label{Ailap}
\end{equation}
\begin{equation}
\hat T_{s}^{(r)}(x) = T^{(L)}_{s}(x)\sum_{i\ge 0}\left(x X_{s,r}^2 T^{(L)}_{s}(x)\right)^i
\label{hTii}
\end{equation}
Each of the $T^{(L)}_{s}(x)$ is the generating function of primitive
tree walks with root $s$, formed by
trees isomorphic to trees with root $j$, generated by
$T^{(L)}_{j}(x)$ (notice that the trees rooted in $s$ do not contain the vertex $r$
otherwise $B^{(L)}_{r,s}$ would not consist of primitive walks);
each of these $T^{(L)}_{s}(x)$ has different internal edges due to the $abab$
exclusion rule:
if there were an edge corresponding to $X_b$ in common between  two trees
of two $T^{(L)}_{s}(x)$, one would have the product
$X_{r,s}\cdots X_b \cdots X_{s,r}^2 \cdots X_b\cdots$;
with $X_a = X_{r,s} = X_{s,r}^2$,
we would get the forbidden $abab$ pattern.\\

Each $B_{r,s}(x)$ contains ``internal'' unit random vectors and one ``external'' unit random vector associated to $X_{r,s}$.
The averages can be separated in average on internal and external random vectors.\\

\textsl{Proposition 3}.
In the limit $d  \to \infty$, with $t = \frac{Z}{d}$ fixed, the average over the internal variables of the generating function on primitive walks is expressed in term of the generating
function of the moments:
 \begin{eqnarray}
&&\langle B^{(M)}_{r,s}(x)\rangle_I = \alpha_{r,s} X_{r,s} g^{(M)}(x) \quad , \quad \nonumber\\
&& \texttt{where} \quad M=A \quad\texttt{or}    \quad L \quad \texttt{and}\nonumber\\
&&g^{(A)}(x) = x^2 \,f_A(x) \quad , \quad g^{(L)}(x) = \frac{x}{1 - x f_L(x)} \qquad \qquad
\label{axg}
\end{eqnarray}
\\

\textsl{Proof of  Proposition 3}

For the Adjacency matrix, from Eq.(\ref{Aadj}) one has
\begin{eqnarray}
B^{(A)}_{r,s}(x) &=& \alpha_{r,s} x^2 |{\hat a}_{r,s}> <{\hat a}_{r,s}|T^{(A)}_{s}(x)|{\hat a}_{r,s}> <{\hat a}_{r,s}|=\nonumber\\
&=& \alpha_{r,s} X_{r,s} x^2 F^{(A)}_{r,s}(x)
\label{Brsadj} \qquad
\end{eqnarray}
where
\begin{equation}
F^{(M)}_{r,s}(x) \equiv <{\hat a}_{r,s}|T^{(M)}_{s}(x)|{\hat a}_{r,s}>
\label{Frs}
\end{equation}
For each   tree walk,
with $n$ steps and $k$ distinct edges, contributing to $T^{(M)}_{s}(x)$, by \textsl{Proposition 1}
the average over the internal edges gives $t^k$. After performing the internal
averages, the external edge variable appears only in the form
$<{\hat a}_{r,s}|{\hat a}_{r,s}> = 1$, so
 we can write
\begin{equation}
\langle F^{(M)}_{r,s}(x)\rangle_I = f_M(x) 
\label{eFrs}
\end{equation}
and from Eq.(\ref{Brsadj})
\begin{equation}
\langle B^{(A)}_{r,s}(x)\rangle_I = \alpha_{r,s} X_{r,s} x^2 f_A(x)
\label{Badj}
\end{equation}
The analogous derivation, for the Laplacian matrix, gives
\begin{eqnarray}
\langle B^{(L)}_{r,s}\rangle_I& &= 
\langle \alpha_{r,s}X_{r,s}\Big(x + x^2 F^{(L)}_{r,s} \sum_{i\ge 0} (xF^{(L)}_{r,s} )^i\Big)\rangle_I=\nonumber\\
&=& \alpha_{r,s} X_{r,s} \frac{x}{1 - x f_L(x)}
\label{Blapl}
\end{eqnarray}
where it has been used $\langle (F^{(L)}_{r,s})^i\rangle_I = (\langle F^{(L)}_{r,s}\rangle_I)^i = \left(f_L(x)\right)^i$,
which follows from the fact that the
walks of two $T_s$'s have different internal edges, due to the $abab$
exclusion rule.

\textsl{Proposition 4}. 
In the limit $d\to \infty$, with $t = \frac{Z}{d}$ fixed, the resolvents of the Adjacency and Laplacian 
matrices, are those for the effective medium approximation and the 
Marchenko-Pastur distribution.

\textsl{Proof of Proposition 4}.
In the limit $d\to \infty$, the internal edges of two different primitive
walks are all distinct because of Proposition 1,
so that the average over the internal edges
of a product of functional generators $B_{r,s}$ factorizes;
hence from Eqs.(\ref{axg},\ref{T00A})
 \begin{eqnarray}
d\,f(x)&=&\langle tr T_{j_0}(x) \rangle=\sum_{n \geq 0}\langle\langle tr \left( \sum_j   B_{j_0,j}(x)\right)^n\rangle_I\rangle_E=\nonumber\\
&=&\langle tr \sum_{n \geq 0}   \left( g(x) \, \sum_j \alpha_{j_0,j}X_{j_0,j} \right)^n\rangle_E
\quad \label{pro.4}
\end{eqnarray}
 Using Eq.(\ref{p.1})
\begin{eqnarray}
f(x)=1+\sum_{n \geq 1}P_n(t)\,\left( g(x)\right)^n
 \nonumber
\end{eqnarray}
Using the generating function for the Narayana polynomials \cite{peter}
one gets
  %\onecolumngrid
\begin{widetext}

  \begin{equation}
f(x) = \Big(1 + g(x)(1-t) - \sqrt{1-2g(x)(t+1) + g^2(x)(t-1)^2}\Big)/(2g(x))
\end{equation}
\end{widetext}

%\twocolumngrid

so that one obtains
\begin{equation}
g(x)\left(f^2(x) - f(x)(1-t)\right) = f(x) -1
\label{rec1}
\end{equation}

By use of Eq.(\ref{axg}) one obtains the algebraic equations for the
Adjacency matrix  and for the Laplacian matrix
\begin{eqnarray}
&& x^2 f_A(x)^3 - x^2 f_A(x)^2(1-t) - f_A(x) + 1 = 0 \quad , \nonumber\\
&&2x f_L(x)^2 + f_L(x)(xt - 1 - 2x) + 1 = 0
\label{sc}
\end{eqnarray}

The resolvent Eq.(\ref{ris}) is $z\,r_M(z)=f_M(x)$ with $z=1/x$. 
Then the first Eq.(\ref{sc}) is a cubic equation for the resolvent of the Adjacency matrix

$$ r^3_A(z)+\frac{t-1}{z} r^2_A(z)-r_A(z)+\frac{1}{z}=0$$
which is the effective medium approximation by Semerjian and Cugliandolo \cite{semer}.

 The second Eq.(\ref{sc}) is a quadratic equation for the resolvent of the Laplacian matrix
$$2z\, r^2_L(z)+(t-2-z)r_L(z)+1=0$$
It corresponds to the Marchenko-Pastur spectral distribution in Eq.(\ref{z.9}).\\

\section{Conclusions}
The moments of the spectral distribution of the sparse random block matrix 
model introduced in \cite{CKMZ} can be studied for any $d$ analyzing
the same class of walks used in the $d=1$ case, which is the Erdos-Renyi random
graph model.
In the latter model the moments are given by recurrence relations in 
\cite{bau}, \cite{khor}, but
the analytic representation of the spectral distribution
is not known.

In this paper we proved that,
in the limit $d \to \infty$ with $\frac{Z}{d}$ fixed,
the nonvanishing contributions to the moments correspond to
the walks, whose sequence of edges form a noncrossing partition.
Due to the simpler structure of these contributions
in this limit, the resolvents
for the Adjacency matrix and the Laplacian matrix can be computed
analytically, and give respectively the Semerjian-Cugliandolo and the
Marchenko-Pastur distributions, as conjectured in \cite{CKMZ}.
We consider this to be the main result of this paper.
\\

Furthermore, several moments of the two spectral distributions
were analytically evaluated for every $Z$ and $d$, which may 
be useful for approximate evaluations of the spectral functions.\\

\section{Acknowledgments}
One of us (G. M. C.) thanks Alessio Zaccone for introducing him to the  sparse random block model analyzed in this paper and Giorgio Parisi for encouraging an automated evaluation of the moments.\\

\appendix
\section{Derivation of the first three moments in the Laplacian matrix}

We derive the first three moments of  the limiting Laplacian matrix.
At the first three orders there are only noncrossing partitions,
so by Proposition 1 the contribution to a moment of a product of $X$ blocks,
$m$ of which are distinct, is $t^m$.
Furthermore the expansion in Eqs.(\ref{Ailap}, \ref{hTii}) holds;
we verify them here  for the laplacian matrix through order $x^3$, by
comparing Eq.(\ref{T0lapl}) with the primitive walk decomposition
Eq.(\ref{T00A}), which at the first three orders reads
\onecolumngrid
\begin{eqnarray}
T_{j_0}^{(L)} =& 1 + \sum_{j_1} \Big(B_{j_0,j_1} + (B_{j_0,j_1})^2 + 
\sum_{j_2}B_{j_0,j_1} B_{j_0,j_2} + 
B_{j_0,j_1}B_{j_0,j_1}(B_{j_0,j_1} + \sum_{j_2} B_{j_0,j_2}) + \nonumber\\
&\sum_{j_2}B_{j_0,j_1}B_{j_0,j_2}(B_{j_0,j_1} + B_{j_0,j_2}+\sum_{j_3}B_{j_0,j_3})
\Big) + O(x^4)
\label{T00B}
\end{eqnarray}
where here and in the following the sums are over $j_r \neq j_s$ for $r \neq s$.

We write for short $(\alpha X)_{i,j} = \alpha_{i,j}X_{i,j}$, and
$(\alpha X)^2_{i,j} = \alpha_{i,j}X_{i,j}X_{j,i}$ for a diagonal term.

From Eq.(\ref{T0lapl})
\begin{equation}
T_{j_0}^{(L)}(x) = 1 + x L_{j_0,j_0} + x^2 (L^2)_{j_0,j_0} +
    x^3 (L^3)_{j_0,j_0} + O(x^4)
\label{Tj0}
\end{equation}
Denote by $[x^k]f$ the term $x^k$ of the series in $x$ of $f$.

One has
\begin{equation}
[x]T_{j_0}^{(L)} = \sum_{j_1} (\alpha X)_{j_0,j_1}^2
\label{T01}
\end{equation}

\begin{equation}
\nu_1 = \lim_{N\to\infty} \frac{1}{Nd} \sum_{j_0}<tr ([x]T_{j_0}^{(L)})> = 
\lim_{N\to\infty} \frac{1}{Nd} N(N-1)<tr (\alpha X)_1> = \frac{Z}{d} = t
\end{equation}
since all the terms in the sums over $j_0,j_1$ give the same contribution.

One can compute order by order $(L^n)_{j_0,J}$ using the following
formula: let $P_S$ be a product of $(\alpha X)$'s with a set $S$ of indices,
all different. In $P_S \sum_{J_1} L_{k, J_1} L_{J_1,J}$ separate $J_1$
in one of the indices in $S-\{k\}$ or $k$ or a new index $j'$;
one of the resulting
terms contains $L_{k,k} = \sum_{J_2} (\alpha X)_{k,J_2}^2$;
separating $J_2$ in the indices present on in a new index, we get

\begin{eqnarray}
P_S \sum_{J_1} L_{k, J_1} L_{J_1,J} =
\sum_{h \in S-\{k\}} P_S\bigg(-(\alpha X)_{k,h} L_{h,J} + (\alpha X)_{k,h}^2 L_{k,J}\bigg) + \nonumber\\
+ \sum_{j'} P_S\bigg(-(\alpha X)_{k,j'} L_{j',J} + (\alpha X)_{k,j'}^2 L_{k,J}\bigg)
\label{PLL}
\end{eqnarray}

where $j'$ is a new index, not in $S+\{k\}$, and where a term of the sum on
$h$ contributes only if no loop in the associated graph is formed.
For $P_S=1$ and $k=j_0$ one gets

\begin{eqnarray}
(L^2)_{j_0,J} = \sum_{j_1} \bigg((\alpha X)_{j_0,j_1}^2L_{j_0,J} - (\alpha X)_{j_0,j_1}L_{j_1,J}\bigg)
\label{L2J}
\end{eqnarray}
Taking in this equation $J=j_0$ and expanding $L_{j_0,j_0}$ one gets
\begin{equation}
[x^2] T_{j_0}^{(L)} = \sum_{j_1}\big( (\alpha X)_{j_0,j_1}^2 (\alpha X)_{j_0,j_1}^2 + (\alpha X)_{j_0,j_1} (\alpha X)_{j_1,j_0}
+  \sum_{j_2} (\alpha X)_{j_0,j_1}^2 (\alpha X)_{j_0,j_2}^2 \big)
\label{T02}
\end{equation}
from which
\begin{equation}
\nu_2 = \lim_{N\to \infty}\frac{1}{Nd}\sum_{j_0}<tr ([x^2] T_{j_0}^{(L)})> = 2t + t^2
\end{equation}

From Eqs.(\ref{T01}, \ref{T02}, \ref{T00B}) one gets
\begin{equation}
B_{j_0,j_1} = x (\alpha X)_{j_0,j_1}^2 + x^2 (\alpha X)_{j_0,j_1} (\alpha X)_{j_1,j_0} + O(x^3)
\label{Bx2}
\end{equation}
        %\twocolumngrid

which agrees Eqs.(\ref{Ailap},\ref{hTii}) with $\hat T_{j_1}^{(j_0)} = 1 + O(x)$.

Expand $(L^3)_{j_0,J} = \sum_{J_1} (L^2)_{j_0,J_1}L_{J_1,J}$ using
Eqs.(\ref{PLL},\ref{L2J})
\begin{eqnarray}
(L^3)_{j_0,J} = \sum_{j_1}
\Big(
(\alpha X)_{j_0,j_1}(\alpha X)_{j_1,j_0} +
(\alpha X)_{j_0,j_1}^2
\big((\alpha X)_{j_0,j_1}^2 + \sum_{j_2}(\alpha X)_{j_0,j_2}^2\big) 
\Big) L_{j_0,J}-\nonumber\\
\Big((\alpha X)_{j_0,j_1}^2 (\alpha X)_{j_0,j_1} +
(\alpha X)_{j_0,j_1}\big((\alpha X)_{j_1,j_0}^2 +
\sum_{j_2} (\alpha X)_{j_1,j_2}^2\big)
\Big)L_{j_1,J} + \nonumber\\
\sum_{j_2}\big(-(\alpha X)_{j_0,j_1}^2 (\alpha X)_{j_0,j_2} + 
(\alpha X)_{j_0,j_1}(\alpha X)_{j_1,j_2}
\big)L_{j_2,J}
\label{L3}
\end{eqnarray}
Then expand this expression for $J=j_0$ to get $[x^3]T_{j_0}^{(L)}$;
the last term in Eq.(\ref{L3}) becomes
$-\sum_{j_1,j_2} (\alpha X)_{j_0,j_1} (\alpha X)_{j_1,j_2} (\alpha X)_{j_2,j_0}$,
which gives vanishing contribution to $\nu_3$, since the corresponding
graph is a loop.

We separate the contributions to $[x^3]T_{j_0}^{(L)}$ of the various $\prod B$ terms in
Eq.(\ref{T00B}) and match them with the expression for $[x^3]T_{j_0}^{(L)}$
obtained above:
%\onecolumngrid

\begin{equation}
[x^3]\sum_{j_1}B_{j_0,j_1} = \sum_{j_1} (\alpha X)_{j_0,j_1}\big((\alpha X)_{j_1,j_0}^2 + \sum_{j_2}(\alpha X)_{j_1,j_2}^2\big)(\alpha X)_{j_1,j_0}
\label{T3B1}
\end{equation}
\begin{equation}
[x^3]\sum_{j_1}(B_{j_0,j_1})^2 = \sum_{j_1} (\alpha X)_{j_0,j_1}(\alpha X)_{j_1,j_0}(\alpha X)_{j_0,j_1}^2 + (\alpha X)_{j_0,j_1}^2 (\alpha X)_{j_0,j_1}(\alpha X)_{j_1,j_0}
\label{T3B2}
\end{equation}
\begin{equation}
[x^3]\sum_{j_1,j_2}B_{j_0,j_1}B_{j_0,j_2} = \sum_{j_1,j_2 }(\alpha X)_{j_0,j_1}(\alpha X)_{j_1,j_0}(\alpha X)_{j_0,j_2}^2 + (\alpha X)_{j_0,j_1}^2(\alpha X)_{j_0,j_2}(\alpha X)_{j_2,j_0}
\label{T3B3}
\end{equation}
\begin{eqnarray}
[x^3](\sum_J B_{j_0,J})^3 &= \sum_{j_1}((\alpha X)_{j_0,j_1}^2)^3 + \sum_{j_1,j_2}\Big(((\alpha X)_{j_0,j_1}^2)^2(\alpha X)_{j_0,j_2}^2 + \nonumber\\
& (\alpha X)_{j_0,j_1}^2(\alpha X)_{j_0,j_2}^2\big((\alpha X)_{j_0,j_1}^2 + (\alpha X)_{j_0,j_2}^2 + \sum_{j_3}(\alpha X)_{j_0,j_3}^2\big)\Big)
\label{T3B4}
\end{eqnarray}

        %\twocolumngrid
so that $\nu_3 = 4t + 6t^2 + t^3$.\\

From Eqs.(\ref{Ailap}, \ref{T3B1}) we get
\begin{equation}
\hat T_{j_0}^{(j_1)}(x) = 1 + x\sum_{j_2}(\alpha X)_{j_1,j_2}^2 + x(\alpha X)_{j_1,j_0}^2 + O(x^2)
\label{T3B1a}
\end{equation}
which satisfies Eq.(\ref{hTii}).

Eqs.(\ref{T3B2},\ref{T3B3}, \ref{T3B4}) follow trivially from Eq.(\ref{Bx2}).

\section{ The  moments in the Adjacency and Laplacian matrices, for generic $d$.}

 We report here  the analytic evaluation of several  moments, helped by computer symbolic enumeration. In the case of the Adjacency matrix, we evaluated the moments up to $\mu_{26}=\lim_{N \to \infty}\frac{1}{N\,d} <{\rm Tr}\,A^{26}>$.\\

In the case of the Laplacian matrix, we evaluated the moments up to $\nu_{15}=\lim_{N \to \infty}\frac{1}{N\,d} <{\rm Tr}\,L^{15}>$.

 We report here \cite{per} the moments only up to $\mu_{18}$ and $\nu_{10}$ . The moments are displayed in terms of the two variables $t=Z/d$ and $d$, but the latter variable only appears in  the coefficients $c_m$, defined in Eq.(\ref{cdef}).  
 This representation is useful to read the spectral moments for fixed $t=Z/d$ and the extreme values of $d$ :  $d=1$ and $d=\infty$. Indeed every $c_m=1$ for $d=1$ and every $c_m=0$ for $d=\infty$.\\
We compared these moments in the case $d=1$ using the recurrence equations
\cite{bau}, \cite{khor}; in the case $d=\infty$ using the resolvents
of the Marchenko-Pastur and Semerjian-Cugliandolo distributions.
We have not checked our results for the moments with direct numerical
simulations with random matrices.

Let us note that the form taken by the averages depends on the order
in which the variables are integrated; the results are given for a
particular choice of order in the averages, so they could be written in
other equivalent forms.

%\onecolumngrid
\begin{widetext}

  \begin{eqnarray} 
        \mu_k&=&\lim_{N \to \infty}\frac{1}{N\,d} <{\rm Tr}\,A^k>  \quad , \quad \mu_0=1\quad , \quad   \mu_{2k+1}=0  \nonumber\\
 \mu_2 &=& t  \qquad , \qquad  \mu_4 = t+ 2 t^2 \qquad , \qquad
 \mu_6 = t + 6t^2 +5t^3 \nonumber\\
\mu_8 &=&t+ t^2   \left( 12 +2\, \,c_2\right)  +28 \, t^3+14\,t^4  \nonumber\\
 \mu_{10} &=& t+t^2\left( 20+10\,c_2\right)
         +t^3\left( 90+20\,c_2\right) +
         120\, t^4 +42\,t^5 \nonumber\\
        \mu_{12} &=& t+t^2\left(30+30 \,c_2+2\,c_3\right)+
 t^3 \left( 220+ \frac{5}{3}c_2 (88+5\,c_2) \right) +\nonumber\\
&+& t^4 \, \left(550+132\,c_2\right)+ 495 t^5 +  132 t^6 \nonumber\\
\mu_{14} &=& t+t^2 \left(42+ 70\,c_2+14\,c_3\right)+t^3\left( 455+\frac{c_2}{3}\left(1820+301\,c_2\right)+28\,c_3
 \right)+\nonumber\\
&+&t^4\left(1820+  \frac{c_2}{3}(3934+350\,c_2\right)+
t^5\left(3003+728\,c_2\right)   +2002 t^6 + 429 t^7 \nonumber\\
\mu_{16} &=& t+t^2\left(56+140\,c_2+56\,c_3+2\,c_4\right)+
\nonumber\\
&+&t^3\left( 840+160 \frac{c_2}{3} \left(35+c_3\right)+256\,c_3+612\,(c_2)^2\right) +\nonumber\\
&+&t^4\left( 4900+2 \frac{c_2}{3}\left(10597+74\,(c_2)^2\right)+1808\,(c_2)^2+240\,c_3\right)+\nonumber\\
&+&t^5\left( 12740+9280\,c_2+1000\,(c_2)^2\right)+t^6\left(15288+3640\,c_2\right)+ 8008 t^7 +
1430 t^8 \nonumber\\
  \mu_{18} &=& t+t^2\left(72+252\,c_2+168\,c_3+18\,c_4 \right)+\nonumber\\
&+&t^3\left( 1428+4760\,c_2+\frac{c_3}{5}\left( 103\,c_3+3342\,c_2\right)+2596\,(c_2)^2+1296\,c_3+36\,c_4\right)+\nonumber\\
&+&t^4\left( 11424+27462 \,c_2+4\frac{ (c_2)^2}{3}\left(10372+815\,c_2\right)+960\,c_2\,c_3+2754\,c_3\right)+\nonumber\\
&+&t^5\left( 42840+  62484\,c_2  + 1632\,c_3 + 19185\,(c_2)^2  +888 \,(c_2)^3  \right)+\nonumber\\
&+&t^6\left(79968+ 57256\, c_2  +6800\,(c_2)^2  \right)+t^7\left(74256 +17136\,c_2  \right) +\nonumber\\
&+&  31824 t^8 + 4862 t^9
  \qquad
        \label{z.1}
\end{eqnarray}

        \begin{eqnarray}
  \nu_k &=&\lim_{N \to \infty}\frac{1}{N\,d} <{\rm Tr}\,L^k> \quad , \quad \nu_0=1 \nonumber\\
  \nu_1 &=& t \qquad , \qquad
 \nu_2 = 2\, t + t^2 \qquad , \qquad
\nu_3 = 4\, t +6\, t^2 + t^3  \nonumber\\
\nu_4 &=& 8\, t + t^2 \left( 24 + c_2 \right)+12 t^3 +t^4  \nonumber\\
\nu_5 &=& 16\, t + t^2 \left(80  +10\, c_2\right)+t^3 \left(80 +5\, c_2\right) +
20\,t^4 + t^5 \nonumber\\
\nu_6 &=&32\,  t+ t^2 \left(240+60 \, c_2 +c_3 \right)+ 
t^3 \left(400+72 \, c_2+ 4 \,c_2 \frac{ 1+2\, c_2}{3} \right) +\nonumber\\
&+& t^4 \left( 200+15\, c_2  \right)+30 t^5 +t^6\nonumber\\
\nu_7 &=&64\,  t+ t^2 \left(672+280\, c_2  +14\, c_3 \right)+\nonumber\\
&+& t^3 \left[1680+588\, c_2  +14\left( c_2 \right)^2+7\, c_3  +
 56\frac{c_2}{3}\left( 1+2\, c_2\right)\right]+\nonumber\\
&+& t^4 \left[ 1400+294\, c_2  +28 \frac{c_2}{3}\left(1+2\,c_2 \right)\right]+
 t^5 \left( 420+35\, c_2  \right)+ 42\,t^6+t^7 
 \nonumber\\
 \nu_8 &=& 128\,  t+ t^2 \left( 1792+1120\, c_{2} + 112\, c_{3} + c_{4} \right)+\nonumber\\
&+&t^3 \left( 6272+  128\, c_{3}+56 \frac{c_2}{3}\left(c_3+200\right)+544 (c_{2})^{2}   \right)+\nonumber\\
&+&t^4 \left( 7840+28\, c_{3}+\frac{c_2}{3} ( 9925+1412\,c_2)+ 12 (c_{2})^{3}     \right)+\nonumber\\
&+&t^5 \left(3920+ 112 \frac{ c_{2}}{3} (25+2\, c_2)  \right)+
t^6 \left(784+ 70 c_{2} \right)+ 56 t^7 +t^8 \nonumber\\
 \nu_9 &=& 256\,  t+ t^2 \left(4608+ 4032 c_{2} + 672 c_{3} + 18 c_{4}\right)+\nonumber\\
&+&t^3 \left(21504+ 32\,c_2 (595+131\,c_2)+1296 c_{3} + 4\,\frac{c_3}{5}(11\,c_3+504\,c_2) +  9 c_{4}  \right) +\nonumber\\
&+&t^4 \left(37632+4 \frac{ (c_{2})^{2}}{3} \left(4727+286\,c_2\right) + 168 c_{2} c_{3} + 25950 c_{2} + 648 c_{3} \right)+\nonumber\\
&+&t^5 \left(28224+108 (c_{2})^{3} + 2388 (c_{2})^{2} + 12975 c_{2} + 84 c_{3} \right)+\nonumber\\
&+&t^6 \left( 9408+ 2380\, c_{2} +224 (c_{2})^{2}  \right)+t^7 \left(1344+126 c_{2}\right)+
72 t^8 +t^9 \nonumber\\
\nu_{10} &=& 512 \, t+ t^2 \left(11520+ 13440 \,c_{2} + 3360\, c_{3} + 180\, c_{4} + c_{5}\right) +\nonumber\\
&+&t^3 \left( 69120+26240 c_{2}^{2} +4\frac{c_2}{3}(25\, c_4+3584\,c_3)+
    85120 c_{2} + 248 c_{3}^{2} + 9600 c_{3} + 200 c_{4} \right)+\nonumber\\
&+&t^4 \left(    161280+        20 \frac{(c_2)^2}{9} ( 26897+ 2792\,c_2+137\,c_3)+  4447 \,c_{2} c_{3} + 164300\, c_{2}+\right.   \nonumber\\
&+ & \left. 88\,( c_{3})^{2} + 8100\, c_{3} + 45\, c_{4}  \right)  +\nonumber\\
&+& t^5 \left(169344+ 2  \frac{c_2}{27} \left(1705527+ 56807 (c_2)^2+ 1036 (c_2)^3  \right)   + 39438\, c_{2}^{2} + 840\, c_{2} c_{3} +\right. \nonumber\\
&+ &    \left. 2400\, c_{3}   \right) +\nonumber\\
&+& t^6 \left( 84672+ 540 c_{2}^{3} + 8860 c_{2}^{2} + 41075 c_{2} + 210 c_{3} \right)+t^7 \left(   560 c_{2}^{2} + 5320 c_{2} + 20160 \right) +\nonumber\\
&+& t^8 \left( 2160+210 c_{2} \right)+90 t^9 +t^{10} 
\nonumber\\
        \label{z.2}
\end{eqnarray}
%\twocolumngrid
\end{widetext}

\section{Spectral density of $L_{1,1}$.}

Every diagonal block of the Laplacian matrix is the sum of $(N-1)$ identically distributed random matrices.
$$L_{1,1}=\sum_{j=2}^N \alpha_{1,j} X_{1,j}$$

In the $d=1$ case, each block $X_{1,j}$ is replaced by one and the
 probability law of a diagonal entry of $L$ for large $N$, is the Poisson distribution of parameter $Z$
$$P\left( L_{1,1}=\sum_{j=2}^N \alpha_{1,j}=k\right)= \frac{Z^k}{k!} e^{-Z}$$

The moments $m_s$ of Poisson distribution are
\begin{eqnarray}
m_s &=&\lim_{N \to \infty} < \left(\sum_{j=2}^N \alpha_{1,j}\right)^s>=\nonumber\\
 &=&\sum_{k=0}^\infty k^s\frac{Z^k}{k!} e^{-Z}=\sum_{i=1}^s Z^i  \,S(s,i)
\nonumber
  \end{eqnarray}

where $ S(s,i)$ are the Stirling numbers of second kind, that is the number of partitions of a $s$-set $\{X\}=\{X_1, X_2, \dots , X_s\}$ into $i$ parts. \\

 For generic dimension, $1<d<\infty$,
we  computed the moments through $m_{15}$. We report here the first five.

%\onecolumngrid
\begin{widetext}

 \begin{eqnarray}
m_s&=&\frac{1}{d} \lim_{N\to \infty}< \texttt{tr}\left( \sum_{j=2}^N \alpha_{1,j} X_{1,j} \right)^s>
 \qquad , \qquad m_1= t \equiv Z/d \nonumber\\
 m_2 &=& t + t^2 \nonumber\\
m_3 &=& t +3 t^2+t^3 \nonumber\\
m_4 &=& t + t^2\left(6+c_2\right)+6 t^3+t^4 \nonumber\\
m_5 &=&  t + t^2 \left(10+5\,c_2\right)+t^3 \left(20+5\,c_2\right)+10\,t^4 +t^5 \qquad \qquad
        \label{z.6}
\end{eqnarray}
%\twocolumngrid
\end{widetext}

According to a theorem by L. Pastur \cite{book} the spectral density of the matrix $L_{1,1}$ in the limit $N \to \infty$ , $d \to \infty$ with the ratio $\frac{N}{d}$ fixed , is the spectral density
 \begin{eqnarray}
\rho_{MP}(\lambda)&=&\frac{ \sqrt{(a_{+}-\lambda)(\lambda-a_{-})}}{2\pi \lambda } \quad,\nonumber\\
  a_{\pm}&=&(1 \pm \sqrt{t})^2
        \label{z.7}
  \end{eqnarray}
        The moments $P_s(t)=\int_{a_{-}}^{a_{+}} \lambda^s \rho_{MP}(\lambda)\,d\lambda$ are the Narayana polynomials
$$P_s(t)=\sum_{j=1}^s N(s,j)\,t^j$$
In the Section IV \textsl{Proposition 2}, this result is derived by a combinatorial argument; the difference with
respect to the $d=1$ case is that the restriction to the terms without
the $abab$ pattern reduce the number of partitions of $n$ elements with $k$
blocks from $S(n,k)$ to $N(n,k)$.

%\end{appendix}

\twocolumngrid


\begin{thebibliography}{}
\bibitem{per} A reader interested in the moments  of higher order, not quoted here, is welcome to write to M. Pernici.
\end{thebibliography}
\end{document}